\documentstyle[aps,floats,twocolumn,graphicx]{revtex}
\begin{document}
\draft
\wideabs{
\title{Possible scenario of the spatially separated Fermi-Bose mixture in
the superconductive bismuthates Ba$_{1-x}$K$_x$BiO$_3$}
\author{M.~Yu. Kagan$^{\mathit a,b}$}
\address{$^a$P.~L. Kapitza Institute  for Physical Problems, Kosygin  Str. 2,
117334  Moscow, Russia.}
\address{$^b$Max-Planck-Institut  f\"{u}er Physik Komplexer Systeme,
N\"{o}thnitzer Str. 38, D-01187, Dresden, Germany.}
\author{A.~P. Menushenkov, K.~V.  Klementev, and A.~V. Kuznetsov}
\address{Moscow State Engineering Physics
Institute, Kashirskoe shosse 31, 115409 Moscow, Russia.}

\date{\today}
\maketitle
\begin{abstract}
A new scenario for superconductivity in the bismuthates based on the concept of
a spatially separated Fermi-Bose mixture is proposed. In the framework of this
scenario we qualitatively explain the insulator-metal transition and the nature
of the superconductivity which occur in Ba$_{1-x}$K$_x$BiO$_3$ under doping by
K. We also analyze briefly an applicability of this scenario for the high-$T_c$
superconductors. In conclusion some additional experiments required to
elucidate more precisely the nature of the superconductivity in the bismuthates
are discussed.  \end{abstract} }

\section{Introduction}\label{Intr}
The concept of the extremely strong-coupling superconductivity with the
preexisted local pairs was firstly introduced by Shafroth \cite{Shafroth} in
the middle of fifties. His statement was that in the extremely type-II
superconductors, where the parameter $\xi_0k_F\lesssim1$, the nature of the
superconductive transition corresponds to the local pair formation (pairing in
the real, and not in the momentum space) at some relatively high temperature
$T^*$, and their Bose-Einstein condensation (BEC) at a lower critical
temperature $T_c<T^*$. Later on Alexandrov and Ranninger developed this
concept \cite{Alexandrov} in connection with the narrow-band materials with an
extremely strong electron-phonon coupling constant ($\lambda\gg1$), where a
standard Eliashberg theory \cite{Eliashberg} starts to fail. The key issue of
their approach was a statement that in narrow bands, where polaron formation is
important, it is possible, in principle, to create the conditions where the two
polarons could effectively occupy the same potential well, prepared in a
self-consistent fashion. Approximately in the same time Leggett and Nozieres
\cite{Leggett,Nozieres} developed a general theory which yields a smooth
interpolation between a BCS-type of pairing in the momentum space for a small
electron-electron attraction and the pairing in the real space for a large
electron-electron attraction.

There were two crucial points in the papers \cite{Leggett,Nozieres}. (i) Their
results are valid independently of the precise nature of the short-range
effective attraction between electrons; (ii) they investigated
self-consistently a standard equation for the superconductive gap and an
equation for the number of particles conservation. The most important result of
Nozieres and Leggett is that for $T^*>T_c$ (or, in other words, for a binding
energy of a local pair $|E_b|>\varepsilon_F$) the chemical potential $\mu$ is
always large and negative $\mu=-|E_b|/2+\varepsilon_F<0$. Hence in a
strong-coupling limit a superconductive pairing takes place not on the
Fermi-surface, as in the BCS-theory, but below the bottom of a conductive band.
This is a crucial drawback of all local-pairs theories.  We cannot match two
basic facts: the existence of the Fermi-surface and the presence of preformed
pairs above $T_c$. First one who emphasized this contradiction was J.~Ranninger
\cite{Ranninger,Chakraverty}, who introduced the concept of the two-band
Fermi-Bose mixture. In this scenario the presence of a degenerate fermionic
band guarantees the existence of a Fermi-surface, while the Bose-Einstein
condensation is responsible for the superconductivity in a bosonic band.  Soon
after the discovery of high-$T_c$ superconductors P.~W. Anderson
\cite{Anderson} reintroduced the concept of the local pairs.  He also
introduced two bands of the fermionic and bosonic quasiparticles.  In his
approach, a superconductive transition was connected with the BEC in the
bosonic band of the charge excitations --- holons, while the presence of a
large Fermi-surface was guaranteed by the fermionic band of the spin
excitations --- spinons.  Unfortunately, even this beautiful approach was not
totally successful because at least in a one layer the BEC of holons yields a
charge of a superconductive pair equal to $e$ instead of $2e$ measured
experimentally.  Later on Geshkenbein, Ioffe, and Larkin \cite{Geshkenbein}
phenomenologically reintroduced a concept of the Fermi-Bose mixture on the
level of the coefficients in the Ginzburg-Landau expansion and showed that
several important experiments in the underdoped high-$T_c$ materials can be
explained naturally within this form of the Ginzburg-Landau functional.

So, up to now a question whether a two-band Fermi-Bose mixture scenario is
applicable to the high-temperature superconductive (HTSC) materials is still
open. Probably, the best chances to be described by this scenario has a bismuth
family of high-$T_c$ superconductors Bi$_2$Sr$_2$CaCu$_2$O$_{8+\delta}$, where
the parameter $k_F\xi_0\sim2$, and the tunneling experiments of Fischer {\em et
al.}\cite{Rener} signal the formation of a pretty large and a stable pseudogap
at temperatures well above $T_c$. In our paper we would like to discuss a
possibility of a two-band Fermi-Bose mixture scenario in a quite different
class of superconductors with a relatively high $T_c\sim30$\,K, namely in the
superconductive bismuthates Ba$_{1-x}$K$_x$BiO$_3$ (BKBO). The key issue of our
paper is a possibility of the existence of {\em the two spatially separated
bands} of fermionic and bosonic quasiparticles in these materials.

Our paper is organized as follows. In the first part we present the basic
experimental facts concerning the local electronic and crystal structure
peculiarities, and their connection with the superconductive and the normal
transport properties of BKBO. In the second part we try to show how
these basic facts could be naturally explained within a scenario of the two
spatially separated bands of the fermionic and bosonic quasiparticles.  We
conclude the paper by a summary of our model and a discussion of the several
additional experiments, which would help us to give a definite answer whether
our proposal is the only possibility for a superconductive pairing in the
bismuthates.

\section{Spatially separated Fermi-Bose mixture}\label{mixture}

The BaBiO$_3$, which is a parent compound for the bismuthates
Ba$_{1-x}$K$_x$BiO$_3$ and BaPb$_{1-x}$Bi$_x$O$_3$ (BPBO), represents a
charge-density-wave (CDW) insulator having the two gaps: an optical gap
$E_g=1.9$\,eV and an activation (transport) gap $E_a=0.24$\,eV \cite{Uchida}. A
partial replacing of Ba by K in BKBO causes the decrease of the gaps, and as a
result the insulator-metal transition as well as the superconductivity are
observed at the doping levels $x\gtrsim0.37$. The superconductivity remains up
to the doping level $x=0.5$ corresponding to the solubility limit of K in BKBO
but a maximal critical temperature $T_c\simeq30$~K is achieved for $x\simeq0.4$
\cite{Cava,Pei2}.

\subsection{Local crystal structure peculiarities}

A three-dimensional character of a cubic perovskite-like structure of the
bismuthates differs from a two-dimensional one in the HTSC cuprates.  The
building block in the bismuthates is a BiO$_6$ octahedral complex (analogue of
CuO$_n$ ($n=4,5,6$) in the HTSC-materials). The octahedral complexes are the
most tightly bound items of the structure because of a strong covalence of the
Bi$6s$-O$2p_\sigma$ bonds. According to the crystallographic data \cite{Cox},
the crystal structure of a parent BaBiO$_3$ represents the alternating
arrangement of the expanded and contracted BiO$_6$ octahedra in a barium
lattice, spoken as a ``breathing'' distortion. Such an alternation together
with a static rotation of the octahedra around [110] axis produce the
monoclinic distortion of a cubic lattice. As shown in
Refs.~\cite{MenIndia,Pisma,MenushenP}, to the larger soft octahedron
corresponds the complex BiO$_6$ with the completely filled Bi$6s$-O$2p$
orbitals and to the smaller rigid octahedron corresponds a complex
Bi\underline{L}$^2$O$_6$. Here \underline{L}$^2$ denotes the free level in the
antibonding Bi$6s$-O$2p_{\sigma^*}$ orbital of the smaller octahedral complex.

The K doping of BaBiO$_3$ is equivalent to a hole doping and leads to a partial
replacement of the larger soft octahedra BiO$_6$ by the smaller rigid octahedra
Bi\underline{L}$^2$O$_6$ \cite{MenushenP}. This causes a decrease and finally a
disappearance of the static breathing and tilting distortions, and the lattice
should contract despite a practically equal ionic radii of K$^+$ and Ba$^{2+}$.
As a result, the average structure according to the neutron diffraction data
\cite{Pei} at the doping level $x=0.37$ becomes a simple cubic. However, the
local EXAFS probes \cite{Yacoby,Pisma,MenushenP} showed the essential
difference of the local crystal structure from the average one. We found out
that the oxygen ions belonging to the different BiO$_6$ and
Bi\underline{L}$^2$O$_6$ octahedra vibrate in a double-well potential, while
those having equivalent environment of the two equal Bi\underline{L}$^2$O$_6$
octahedra oscillate in a simple harmonic potential \cite{Pisma,MenushenP}. This
very unusual behavior is in a close connection with a local electronic
structure of BKBO.

\subsection{Local electronic structure}

\begin{figure}[!t]\begin{center}
\includegraphics*[width=\hsize]{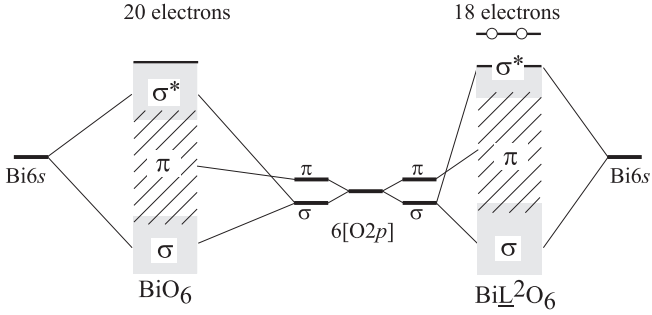}\end{center}
\caption{The scheme of the formation of an electronic structure
for the different octahedral BiO$_6$ complexes.} \label{el_schem}\end{figure}

The coexistence in BaBiO$_3$ of the different types of the octahedra with the
two Bi-O bond lengths and strengths reflects the different electronic
structures of BiO$_6$ complexes. The valence band of BaBiO$_3$ is determined by
the overlap of Bi$6s$ and O$2p$ orbitals \cite{Sleight,Mattheiss}, and, owing
to a strong Bi$6s$-O$2p_\sigma$ hybridization, the octahedra can be considered
as the quasi-molecular complexes \cite{Sugai}. In each complex there are ten
electron levels consisting of the four bonding-antibonding Bi$6s$-O$2p_\sigma$
orbitals and the six nonbonding O$2p_\pi$ orbitals.  A monoclinic unit cell
includes the two octahedra and contains 38 valence electrons (10 from the two
bismuth ions, 4 from the two barium ions, and 24 from the six oxygen ions).
All the Bi-O bond lengths should be equal and the local magnetic moments should
be present in the case of equal electron filling for nearest octahedra
(Bi\underline{L}$^1$O$_6+$Bi\underline{L}$^1$O$_6$). In contrast both the
presence of the two types of octahedral complexes and an absence of any local
magnetic moment are experimentally observed \cite{Uchida,Uemura}, so the
mentioned above scheme of the valence disproportionation
2Bi\underline{L}$^1$O$_6\to$ Bi\underline{L}$^2$O$_6$+BiO$_6$ was proposed
\cite{MenushenP}. In this scheme the numbers of occupied states in the
neighbouring octahedral complexes are different: the octahedron
Bi\underline{L}$^2$O$_6$ contains 18 electrons and has one free level or a hole
pair \underline{L}$^2$ in the upper antibonding Bi$6s$O$2p_{\sigma^*}$ orbital,
while in the octahedron BiO$_6$ with 20 electrons this antibonding orbital is
filled as shown in Fig.~\ref{el_schem}.  It is quite natural that the
Bi\underline{L}$^2$O$_6$ octahedra have the stiff (quasi-molecular) Bi-O bonds
and a smaller radius, while the  BiO$_6$ octahedra represent the non-stable
molecules with a filled antibonding orbital and a larger radius. Because the
sum of the two nearest octahedra radii overcomes the lattice parameter, the
octahedral system must tilt around [110] aixis, producing together with a
breathing a monoclinic distortion in BaBiO$_3$.

Thus, in BaBiO$_3$ one has an alternating system of the two types of the
octahedral complexes filled with the local pairs: the hole pairs in
Bi\underline{L}$^2$O$_6$ complexes and the electron pairs in BiO$_6$ complexes.
The local pair formation in BaBiO$_3$ was advocated previously, see for example
\cite{Uchida,Rice,Varma,DeJongh,Sugai3,Yu,Taraphder}. The binding mechanism for
the pairs is probably of an electronic nature \cite{Varma,Taraphder} in
accordance with Varma's picture of the pairing due to the skipping of the
valence ``4+'' by the Bi ion \cite{Varma}. However one cannot fully exclude the
lattice mediated pairing \cite{Uchida,Rice,Sugai3} in accordance with de
Jongh's statement \cite{DeJongh} that the preference to retain the closed-shell
structures overcomes the Coulomb repulsion involved with an intrasite bipolaron
formation.

\begin{figure*}[!t]\begin{center}
\includegraphics*[width=0.77\hsize]{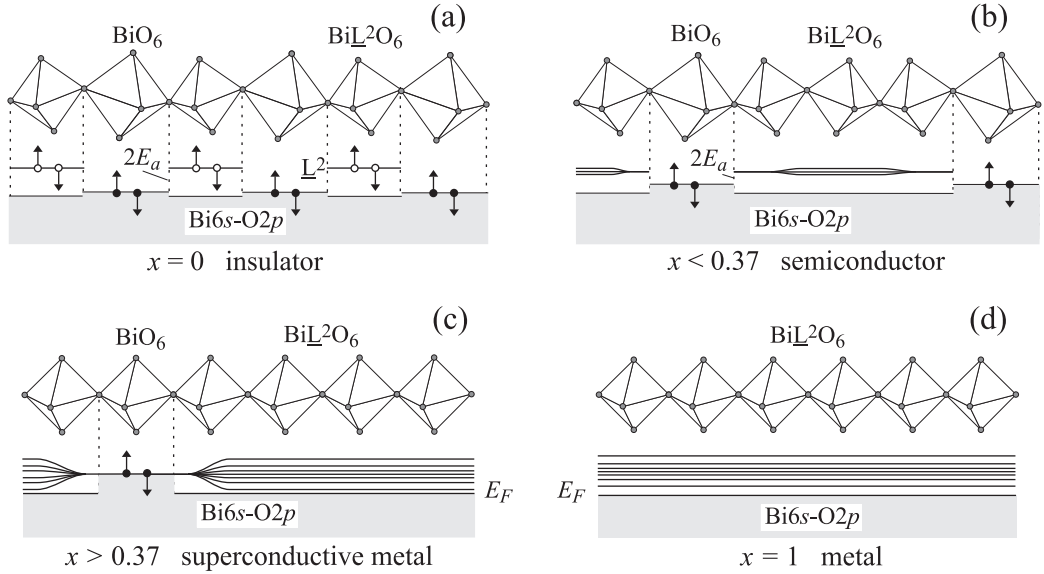}\end{center}
\caption{The scheme of the insulator-metal phase transition for the K-doping of
Ba$_{1-x}$K$_x$BiO$_3$ in the framework of the relationship between the local
crystal and the local electronic structures. The local crystal structure of the
octahedral complexes  (at the top) and the local electronic structure (at the
bottom) are shown on the pictures (a)-(d). The occupied states of the
Bi$6s$O$2p$ valence band are marked by gray. $2E_a$ is the activation gap.
Black and white circles with arrows denote, correspondingly, the electrons and
the holes with the opposite spin orientations. (a) A monoclinic phase of an
insulator BaBiO$_3$. (b) An orthorhombic phase of a semiconducting BKBO at
$0<x<0.37$.  The splitting of free level \underline{L}$^2$ at a spatial overlap
of the Bi\underline{L}$^2$O$_6$ octahedra is sketched. (c) An undistorted cubic
phase of a superconducting metal at $x>0.37$. The formation of a Fermi-liquid
state is shown arising  due to the overlap of an unoccupied fermionic band $F$
with an  occupied Bi$6s$O2$p$ valence band when the percolation threshold is
reached. (d) An undistorted cubic phase of a nonsuperconducting metal at
$x=1$. A Fermi liquid state with Fermi level $E_F$ is shown.}
\label{octahedra}\end{figure*}

\begin{figure*}[!t]\begin{center}
\includegraphics*[width=0.72\hsize]{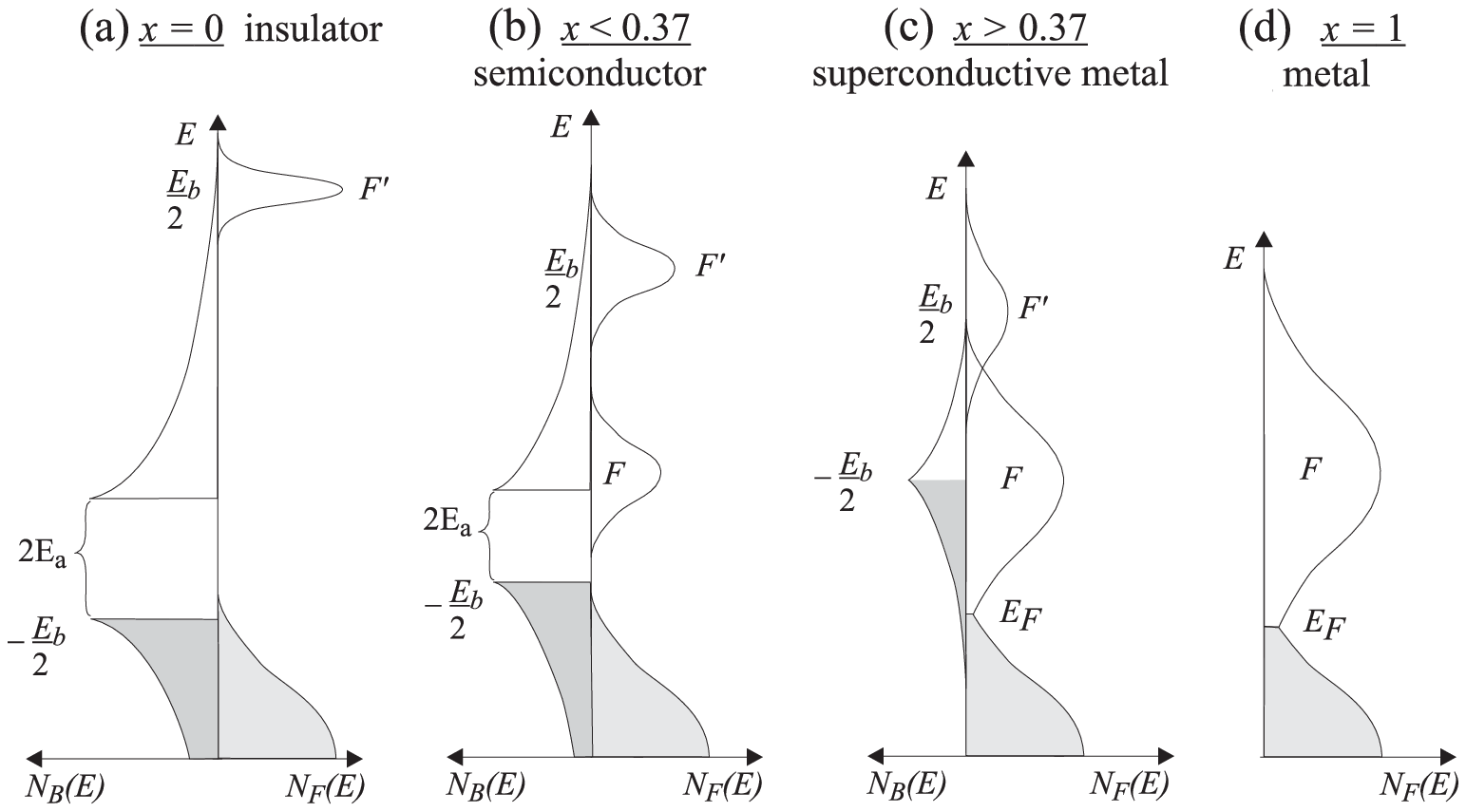}\end{center}
\caption{A sketch of the one-particle density of states for
Ba$_{1-x}$K$_x$BiO$_3$. The contributions from the bosons $N_B(E)$ and the
fermions $N_F(E)$ are depicted separately because the bosonic and fermionic
states are spatially separated. The filled (dark gray) and the unoccupied
(transparent) bosonic bands correspond, respectively, to the contributions from
the electron and the hole pairs. The bands are separated by the activation gap
$2E_a$ which is lowered with a doping level $x$. An empty fermionic band $F'$,
corresponding to the destruction of the pairs, is separated from an  occupied
bosonic band by a binding energy $E_b$. An empty fermionic band $F$ is formed
from an unoccupied bosonic band due to the splitting of the free level
\underline{L}$^2$, which arises from a spatial overlap of the
Bi\underline{L}$^2$O$_6$ octahedra. A filled fermionic band (gray) represents
the Bi$6s$O$2p$ valence band. A band  $F'$ and bosonic bands grow downwards by
the doping because of a decrease of the number of the electron pairs, while a
band $F$  grows due to the increase of the number of free levels. A Fermi
liquid state is formed (c), (d) as a result of the overlap between the band $F$
and the Bi$6s$O$2p$ valence band.} \label{bonds}\end{figure*}

The local electronic structure of BaBiO$_3$ combined with the real-space local
crystal structure is presented in Fig.~\ref{octahedra}(a). In such a system
there are no free Fermi carriers, and the conductivity occurs only due to the
transfer of the carrier pairs \cite{Uchida,MenIndia}. The dissociation of the
pairs and the hopping of a single electron from one octahedron to another, in
similarity with Varma's \cite{Varma} suggestion, cost an energy:  $$E_b=2E({\rm
Bi\underline{L}^1O_6})-[E({\rm BiO_6})+E({\rm Bi\underline{L}^2O_6})],$$
and is observed experimentally as an optical conductivity peak at the photon
energy $h\nu=1.9$\,eV \cite{Uchida}.

Thus we have an example of the normal bosonic liquid of the pairs bound with an
energy $E_b$ (as in Ref. \cite{Kagan}). Experimentally, BaBiO$_3$ shows a
semiconductor-like behavior with an energy gap $E_a=0.24$\,eV, which can be
explained only as a two-particle transport with the activation energy $2E_a$
due to the delocalization of the pairs. From our point of view, the transport
gap is defined by the combined effect of the Coulomb interaction and the
deformation potential between the neighboring octahedral complexes.

In principle, the electronic structure of BaBiO$_3$ presented in
Fig.~\ref{bonds}(a) is similar to that proposed by Namatame {\em et al.}
\cite{Namatame}. They supposed that the thermally excited charge carriers in
BaBiO$_3$ are the polarons with a substantial band narrowing due to a strong
electron-phonon coupling. In our case the carriers are the local pairs (the
bipolarons), the empty and the occupied bosonic bands are very narrow, and that
is why in Fig.~\ref{octahedra}(a) they are shown as energy levels. So, the
transport gap $2E_a$ is the gap between the empty and the occupied bosonic
states, and the conductivity in BaBiO$_3$ arises from the thermo-activated
bosonic quasiparticles. This fact is in agreement with the suggestion of
Taraphder {\em et al.} about a bosonic character of the ground state in
BaBiO$_3$. \cite{Taraphder}

Another key question in the bismuthates is a nature of the optical gap
$E_g=1.9$\,eV. In accordance with the band structure calculations
\cite{Mattheiss}, this gap arises from the peak-to-peak splitting of the Bi$6s$
band due to CDW formation. Namatame {\em et al.} \cite{Namatame} attribute this
gap to the Frank-Condon transition under which the lattice should be frozen.
That contradicts to the observation in the Raman spectra
\cite{Sugai,Sugai3,Tajima3} of the abnormally large amplitude of the
breathing-type vibrations of the oxygen octahedra. The required mode arises
only at a resonant excitation by the laser radiation with an energy $h\nu=E_g$.
The resonance was destroyed and the abrupt decrease of the mode amplitude was
observed when the lasers with an other frequency were used \cite{Tajima3}.
Obviously, the photons cannot excite a carrier pair as a whole and should
destroy it. Such a behavior directly proves that the lattice is involved in
this optical transition. In our scheme an excitation over the optical gap
corresponds to the pair destruction
Bi\underline{L}$^2$O$_6$+BiO$_6\stackrel{h\nu}{\longrightarrow}$
2Bi\underline{L}$^1$O$_6$, which produces a local deformation of the lattice
due to the changing of the two different octahedra on equivalent ones. This
dynamic deformation is manifested in the Raman spectra as an abnormally high
amplitude of the breathing mode. Thus the nature of the optical gap is just the
pair binding energy $E_g=E_b$.  It is important to emphasize that there are no
free fermions in the system. Only the excited fermions can be produced by the
unpairing, and they do not give any input into the charge transport because of
a high value of $E_b=1.9$\,eV. Thus the bosonic and the fermionic subsystems
are separated both spatially and energetically, and therefore the Fermi-Bose
mixture is absent in the parent compound.

\subsection{Formation of the Fermi-Bose mixture}\label{formation}

The substitution of the each two K$^+$ for the two Ba$^{2+}$ modifies the
BiO$_6$ complex to the Bi\underline{L}$^2$O$_6$ one.  As a result, the number
of the small stiff Bi\underline{L}$^2$O$_6$ octahedra increases as
$n_0(1+x)/2$, while the number of large soft BiO$_6$ octahedra decreases as
$n_0(1-x)/2$, where $n_0=1/a^3$ is the number of the unit cells and $a$ is a
lattice parameter. The spatial overlap of the Bi\underline{L}$^2$O$_6$
complexes appears at the finite doping levels, which, taking into account their
small radii and the rigid bonds, contracts the lattice.

The structural changes are accompanied by the essential changes in the local
electronic structure and in the physical properties of BKBO. A spatial overlap
of the \underline{L}$^2$ levels leads to their splitting into an empty
fermionic-like band $F$ inside the
Bi\underline{L}$^2$O$_6-\cdots-$Bi\underline{L}$^2$O$_6$ Fermi-cluster [see
Fig.~\ref{octahedra}(b)]. In the doping range $x<0.37$ the band is narrow
enough due to a polaronic effect and is still separated from an occupied
Bi$6s$O$2p$ subband. The number of the empty electronic states in the $F$ band
increases with $x$ as $\hat{n}_F=n_0(1+x)$, while the number of the local
electron pairs decreases as $n_B=n_0(1-x)/2$.

A free motion of the pairs is still prevented by the intersite Coulomb
repulsion \cite{Varma}, which becomes strongly screened inside the clusters.
When the Fermi-clusters are formed, the conductivity occurs due to the motion
of the pairs through the clusters of the different lengths. The BKBO compounds
demonstrate a semiconducting-like conductivity changing from a simple
activation type to the variable-range-hopping Mott's law \cite{Hellman1}.
Moreover the activation energy lowers with the doping down to $E_a\approx0$ at
$x\approx0.37$.  One can understand the decrease of the activation energy as a
suppression of the Coulomb blockade due to the formation of the Fermi-clusters
and due to the decrease of an energy shift between the empty and the occupied
bosonic levels as the lattice distortion is diminished.

At the doping level $x\approx0.37$ (see Fig.\ref{octahedra}(c) and
Fig.\ref{bonds}(c)) the following cardinal changes take place:

(i) Both the breathing and the rotational static lattice distortions transform
to the dynamic ones. At the cluster borders, where all oxygen ions belongs to
BiO$_6$ and Bi\underline{L}$^2$O$_6$ octahedra,  the local breathing dynamic
distortion is observed as a vibration in a double-well potential of $(1-x)$
part of the oxygen ions \cite{Pisma,MenushenP} but cannot be detected by the
integral methods of the structural analysis such as an X-ray and a neutron
diffraction.

(ii) The infinite percolating Fermi cluster (formed from the spatially
overlapped Bi\underline{L}$^2$O$_6$ octahedra) appears, which leads to the
overlap of an empty fermionic band with a filled one, and as a result $F$
becomes a conduction band.  Overcoming of the percolation threshold provides
the insulator-metal phase transition and the formation of the Fermi-liquid
state for $x>0.37$. The valence electrons of the Bi\underline{L}$^2$O$_6$
complexes previously localized become itinerant which is in agreement with the
experiments \cite{Salem}.

(iii) The pair localization energy disappears $E_a\approx0$ so the local
electron pairs (from the BiO$_6$ complexes) can freely move in the real space
providing a bosonic contribution into the conductivity.  Thus, in the metallic
phase the two types of carriers are present: the itinerant electrons from the
Bi\underline{L}$^2$O$_6$ complexes (fermions) and the delocalized electron
pairs from the BiO$_6$ complexes (bosons). Despite the normal state
conductivity is mainly due to a fermionic subsystem, the contribution from a
bosonic subsystem was also observed by Hellman and Hartford \cite{Hellman} as
the two-particle normal state tunneling.

As a result at doping levels $x>0.37$ we have a new type of a spatially
separated mixture of the bosonic $B$ and the fermionic $F$ subsystems, which
describes metallic properties of BKBO. It should be stressed that the fermions
and the bosons belong to the complexes with the different electronic structure,
therefore {\em the Fermi and the Bose subsystems are spatially separated at any
doping level}.  These subsystems are connected by the relations
$2n_B+\hat{n}_F=2n_0$ and $2n_B/\hat{n}_F=(1-x)/(1+x)$.  The high enough value
of the binding energy, which in the superconductive compositions becomes
apparent as a pseudo-gap $E_b=E_g\approx0.5$\,eV \cite{Blanton}, is the
guarantee against the pair destruction. The unpairing is possible only under
the optical excitation to the band $F'$ (see Fig.~\ref{bonds}), which does not
play any role in the charge transport.

At $x=1$ all the BiO$_6$ octahedra are transformed to the
Bi\underline{L}$^2$O$_6$ ones. The Bose system disappears ($n_B=0$) together
with an excited fermionic band $F'$. As a result, KBiO$_3$ should behave as a
simple Fermi-liquid metal without superconductive properties (see
Fig.\ref{octahedra}(d) and Fig.\ref{bonds}(d)).

It is worth to notice that a metallic KBiO$_3$ compound exists only
hypothetically because of the exceeding of the potassium solubility limit in
BKBO, which in this case equals to $x\approx0.5$.  However BaPbO$_3$ as an
electronic analogue of KBiO$_3$ demonstrates the metallic but not the
superconductive properties. Recent attempts to synthesize KBiO$_3$ at a high
pressure found out that only K$_{1-y}$Bi$_y$BiO$_3$ with a partial replacement
of K$^+$ ions by Bi$^{3+}$ ones is formed \cite{Khasanova}. Such a replacement
should lead to the appearance of the BiO$_6$ octahedra with the local electron
pairs, and hence the compound K$_{1-y}$Bi$_y$BiO$_3$ should be superconductive
in accordance with the discussion above. Indeed, a superconductivity with
$T_c=10.2$\,K was experimentally observed in this compound \cite{Khasanova}.
From this point of view, it follows that BaPbO$_3$ should be superconductive at
a partial substitution of the Ba$^{2+}$ ions for the La$^{3+}$ ones (or for the
other trivalent ions) because such a substitution produces the local electronic
pairs as in the case of K$_{1-y}$Bi$_y$BiO$_3$.

\section{Discussion}\label{Theor}

The most important point which differs our model from the ones considered by
Rice and Sneddon \cite{Rice}, Varma \cite{Varma}, and De~Jongh \cite{DeJongh}
is the claim that the local hole pair belongs to the whole
Bi\underline{L}$^2$O$_6$ complex and not only to the Bi$^{5+}$ site.
Analogously the local electron pair belongs to the whole BiO$_6$ complex and
not only to the Bi$^{3+}$ site. From this point of view we do not believe in
the scheme 2Bi$^{4+}\to$ Bi$^{3+}$+Bi$^{5+}$ of a charge disproportionation,
but propose a scheme 2Bi\underline{L}$^1$O$_6\to$
Bi\underline{L}$^2$O$_6$+BiO$_6$, where \underline{L} denotes a local hole in
the upper antibonding Bi$6s$O$2p_{\sigma^*}$ orbital of the whole octahedral
complex. To some extent our idea has a certain similarity with Zhang-Rice
construction for HTSC-materials \cite{Zhang}.  The essential difference is that
Zhang-Rice singlet is a boson (a holon in Anderson terminology) with zero spin
and charge $e$ (so to create a Cooper pair one needs two singlets). In the
bismuthates from the beginning we have a rather tightly bound boson (biholon)
with a proper charge $2e$. Of course, a total spin of the biholon is again
zero, so their Bose-Einstein condensation corresponds to a standard $s$-wave
superconductivity. The existence of the biholons before was only proved
rigorously in the quasi one-dimensional ladder materials at a strong coupling
along the rungs (Dagotto and Rice \cite{Dagotto}). Our scheme immediately gives
us an understanding of a spatial separation of the fermionic and the bosonic
bands in the bismuthates.

Namely, our model identifies an optical gap with the binding energy
of a preformed pair $E_b=E_g$. In the same time a transport gap $2E_a$
corresponds to the pair localization energy. As a result we have the picture
for the one-particle density of states presented in Fig.~\ref{bonds}. In
this picture for $x=0$ there is a filled bosonic band separated by the large
gap $E_g=E_b$ from an empty fermionic band (an excited band $F'$) and by a
smaller transport gap $2E_a$ from an  empty bosonic band $B$, which plays the
role of the conduction band for the bosonic quasiparticles involved in the
activation transport. In accordance with Ref.~\cite{Kagan} in the
representation of the one-particle density of states the filled bosonic band
has the hole-like dispersion while the empty bosonic band has the electron-like
dispersion. The above bands are on  top of the completely filled wide
($\sim$16\,eV) Bi$6s$-O$2p$ hybridized band, which includes 18 valence
electrons per unit cell. Since the occupied and the empty bands belong to the
different octahedra, they are always spatially separated.

Formally, the transport gap is similar to the impurity band in doped
semiconductors. However, the combined transport and the Hall-effect
measurements showed that the number of carriers participated in an activation
conductivity is extraordinary large and is estimated as a number of unit cells
\cite{Uchida}.  This rejects any impurity mechanisms and suggests that the
conduction process in the doping range $0<x<0.37$ can not be understood by the
conventional transport mechanisms of the ordinary semiconductors. The attempts
have been made to identify the excitation with the energy gap $E_a$ using some
optical methods:  the reflectivity, the  photoconductivity, and the
photoacoustic spectroscopy measurements.  However, no photoresponse could be
detected by either method in the energy region near $E_a$ \cite{Tajima}. So the
meaning of this gap is just a localisation energy of the local pairs. Hence for
$x=0$ we have the normal bosonic semiconductor with an activation character of
the conductivity $\sigma(T)=\sigma_0\exp(-2E_a/kT)$.

As we pointed above, the localization energy is influenced by at least two
effects: the Coulomb intersite repulsion due to CDW state and the energy shift
between the empty and the occupied bosonic bands due to a static lattice
distortion. With the doping both effects diminish, which leads to the decrease
of the pair localization energy. The reason is that the appearing Fermi
clusters screen the Coulomb intersite repulsion and the decrease of the static
lattice distortion lowers the energy shift between the bosonic bands. At
$0<x<0.37$ the lengths of the Fermi-cluster islands along [100] directions are
variable due to the random distribution of the dopant atoms, so the temperature
dependence of the conductivity becomes more complicated, but remains a
semiconducting-like. It corresponds to the experimentally observed
variable-range-hopping conductivity according to the Mott's law
$\sigma=\sigma_0\exp(T_0/T)^{-1/4}$ \cite{Hellman1,Dabrowski}.  The value of
$T_0\simeq(3-6)\times10^8$\,K obtained by Hellman {\em et al.} \cite{Hellman1}
implies a strong carrier localization as in our model. On the language of the
fermionic and the bosonic bands, at $0<x<0.37$ the Fermi and the Bose
subsystems are both energetically and spatially separated.

Near the insulator-metal transition the characteristic temperature $T_0$
depends on the stoichiometry, and an agreement with the Mott's law breaks down
\cite{Hellman1}. A pair localization energy approaches zero because the movement
of BiO$_6$ octahedron, surrounded by the equivalent Bi\underline L$^2$O$_6$
octahedra in undistorted lattice, does not change the total energy. In the
doping range $x\gtrsim0.37$ we have a two-band Fermi-Bose mixture similar to
that proposed earlier by Ranninger {\em et al.} \cite{Ranninger,Chakraverty}.
In the same time this mixture of a bosonic metal state with a Fermi-liquid
state is rather unusual.  The normal conductivity in this mixture is mainly due
to the fermionic subsystem which overcomes at $x\approx0.37$ the percolation
threshold.  Nevertheless, the bosonic two-particle contribution into the normal
state conductivity is also present. At the doping level $x\gtrsim0.37$ the
occupied and the empty pair levels are in the resonance, therefore the energy
separation disappears and the Bose and the Fermi subsystems remain separated
only spatially.

Note that there is an interplay between the Bose and the Fermi subsystems since
the motion of the carrier pairs leads to the transformation of the Bose
octahedral clusters  to the Fermi ones and vice versa in the process of the
dynamic exchange BiO$_6\leftrightarrow \rm Bi$\underline{L}$^2$O$_6$.
Because this process is closely related with the superconductivity, we analyze
it below in more details.

\subsection*{Superconductivity in Ba$_{1-x}$K$_x$BiO$_3$}\label{supercond}

Taking into account an  existence of the double-well potential in
Ba$_{1-x}$K$_x$BiO$_3$ one can consider superconductivity in this compound in
the framework of the anharmonic models for HTSC \cite{Plakida1,Hardy1}. As it
was shown in these models, if the oxygen ions move in a double-well potential,
an order-of-magnitude enhancement of a constant of an electron-lattice coupling
follows automatically from a consistent treatment of this motion. The pairing
mechanism is connected in these models with the  enhancement of the coupling
constant due to the oxygen ion vibrations in a double-well potential. However,
in agreement with Refs.~\cite{Varma,Taraphder}, we believe that in the
bismuthates the pairing mechanism is more probably of the electronic than of
the phonon-mediated  origin (see Sec.~II b). In the same time in our system the
local pairs tunnel between the neighboring octahedra due to the vibration of
the oxygen ions in  the double-well potential.  Therefore we suppose that the
lattice is involved in the superconductivity more probably by providing the
motion of a local pair rather than via a pairing mechanism itself.

\begin{figure}[t]\begin{center}\includegraphics*{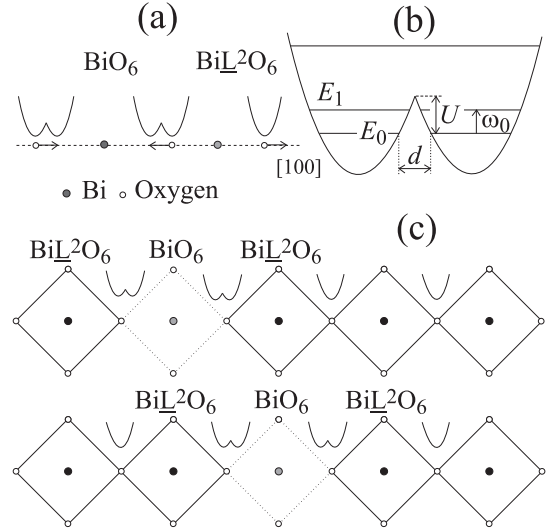}\end{center}
\caption{A sketch of the dynamic exchange
BiO$_6\leftrightarrow \rm Bi$\underline{L}$^2$O$_6$ is shown in the BiO$_2$
plane of the octahedra. (a) A breathing mode of the vibrations along [100]-type
direction of two neighboring octahedra with the different electronic
structures. The BiO$_6$ octahedron transforms to the Bi\underline{L}$^2$O$_6$
one and vice versa due to the electron pair tunneling between the octahedra. An
oxygen ion belonging to such octahedra oscillates in a double-well potential.
An oxygen ion belonging to the equivalent neighboring Bi\underline{L}$^2$O$_6$
octahedra oscillates in a simple parabolic potential.  (b) A double-well
potential with the energy levels for the vibration of the oxygen ion. The
following parameters describe the tunneling barrier between the wells in
Ba$_{0.6}$K$_{0.4}$BiO$_3$ at low temperatures \protect\cite{MenushenP}: the
tunneling frequency $\omega_0\simeq200$\,K, the height of the barrier
$U\simeq500$\,K, the width of the barrier $d\simeq0.07$\,\AA. (c) A motion of
the local electron pair centered on the BiO$_6$ octahedron through the
Bi\underline{L}$^2$O$_6\cdots$ Bi\underline{L}$^2$O$_6$ Fermi cluster. For
detailed explanations see the text.}\label{wells}\end{figure}

The process of a dynamic exchange is illustrated in Fig.~\ref{wells}. An oxygen
belonging to the two neighboring octahedra BiO$_6$ and Bi\underline{L}$^2$O$_6$
vibrates in a double-well potential, and hence the tunneling of the electron
pair between the neighboring octahedra occurs when the ion tunnels through the
potential barrier between the wells.  Because of this interconnection between
the processes of the pair and the oxygen tunneling, one can estimate the matrix
element of the pair tunneling as $t_B\sim\omega_0e^{-D}$ where $\omega_0$ is
the tunneling frequency,
$D=(1/\hbar)\int_{x_0}^{x_1}|p|\,dx\simeq(d/\hbar)\sqrt{2MU}$ is a
quasiclassical transparency of the barrier in the double-well potential, $U$
and $d$ are the barrier height and width, and $M$ is the oxygen ion mass.  Note
that rather small tunneling frequency $\omega_0=200$\,K (see Fig.\ref{wells})
already incorporates the effects connected with a polaronic narrowing of the
one-particle bands $t_p \sim t_0 exp(-g^2)$ and the hopping of a bosonic pair
via virtual dissociation processes. The last ones are described by a second
order perturbation theory (see Ref.~\cite{Nozieres}) and yield an estimate
$t_B\sim t_p^2/|E_b|$ for a width of a bosonic band in a spatially homogeneous
Bose-system.

The tunneling of the pairs helps to establish a macroscopic long range order (a
phase coherence) in the bosonic system.  On the language of the spatially
separated Fermi-Bose mixture, a local pair is transferred from one Bose cluster
to a nearest one over the Fermi-cluster, which, depending on the doping level,
consists of several octahedra. The pairs overcome the Fermi-cluster step by
step. A single step, corresponding to the pair transfer into a neighboring
octahedron, is described by the pair tunneling in the double-well potential.
Thus the tunneling frequency $\omega_0$ is the same for each step. If one
assumes that the steps are independent events, a probability of the overcoming
of the Fermi-cluster can be obtained as a product of the probabilities of an
each step. In this case the matrix element of the pair tunneling through the
Fermi-cluster can be estimated as
$\hat{t}_B\sim\omega_0e^{-\langle N\rangle D}$ where an evarage number of steps
(which is proportional to a Fermi-cluster linear size) can be obtained from
the ratio of the concentrations of Bi\underline{L}$^2$O$_6$ and BiO$_6$
octahedra. Hence a number of steps can be estimated as $\langle
N\rangle\simeq\left(\frac{1+x}{1-x}\right)^{1/3}$.

Of course, it is naturally to assume that the critical temperature of
superconductivity is of the order of the temperature of the Bose--Einstein
condensation $T_c\sim\hat{t}_Ba^2n_B^{2/3}$ in the bosonic system with a large
effective mass $m_B\sim1/\hat{t}_Ba^2$.  We remind that $a^3n_B=(1-x)/2$ in our
case. For $x=0.4$ and the parameters of the double-well potential obtained in
the Ref.~\cite{MenushenP} (see also Fig.~\ref{wells}) we estimated
$T_c\sim50$\,K. This value is larger than the measured $T_c\simeq30$
\cite{Pei2} in the bismuthates.

Note that the estimate above is quite rough.  An accurate analysis of the
superconductivity in the bismuthates should also take into account a
significant boson-phonon interaction, arising due to the interconnection
between the vibrations of the oxygen ions and the transfer of the pairs. When
the pair is transferred from one octahedron to another, the lattice has a
sufficient time to relax, forming each time a new configuration before the next
tunneling event occurs. As a result the pair's ``deformed'' environment (the
BiO$_6$ octahedra) may follow the tunneling processes without the retardation.
If the local pair motion is slow compared to the frequencies of the optical
phonons associated with the deformations of the octahedra, the so-called {\em
anti-adiabatic limit} is fulfilled in our system \cite{DeJongh}.

Note that in our case the stretching phonons are associated with the tunneling
of the pairs. The following conclusions can be made from the analysis of the
phonon modes studied in Ba$_{1-x}$K$_x$BiO$_3$ by an inelastic neutron
scattering \cite{Braden2}. (i) The frequencies of the optical modes of the Bi-O
vibrations $\omega_{ph}>630$\,K are high enough to provide the {\em
anti-adiabatic limit}, since the tunneling frequencies are lower,
$\omega_0\leq300$\,K for $x\leq0.5$ \cite{MenushenP}.  Note that from a
theoretical point of view a polaronic narrowing of the one- particle bandwidths
$t_p<<t_0$, and hence a small value of $\omega_0$ itself, is just a direct
consequence of the non-adiabaticity. (ii) The energy of the longitudinal
stretching mode with [100] wave-vector direction is lower than the energy of
modes with other wave-vector directions or of transversal modes. (iii) The
breathing-type vibrations with the wave vector $\bbox{\rm q}_b$=$(\pi/2a,0,0)$
are energetically favorable since an energy of the longitudinal stretching
phonons is the lowest at the Brillouin band edge. That is why a ``breathing''
of each octahedron should be coordinated with its neighbors to guarantee a
resonant tunneling in the system.  Hence a long-range correlation of the
vibrations should occur at low temperatures when only the low-energy states are
occupied.  (iv) The bandwidth of the longitudinal stretching mode is of the
order of 100\,K, and thus a temperature $T\sim T_c$ is high enough to excite
the non-breathing-type longitudinal stretching phonons with the wave vectors
shorter than $\bbox{\rm q}_b$. The thermal excitation of the phonons with such
short wave-vectors leads to the destruction of the long-range correlation
between the breathing-type vibrations, and hence play a destructive role for
the pair tunneling.

It should be stressed in addition, that the oscillations of the oxygen ion in
the double-well potential corresponds to the breathing-type longitudinal
vibrations with [100] directions (see Fig.~\ref{wells}a).

Thus the pair motion is more correctly described as follows. When the local
pair is transferred to a neighboring octahedron (for example, from the left to
the right in Fig.~\ref{wells}c) due to the transition of the oxygen ion from
one well to another one in the double-well potential, the charge density
corresponding to a pair is quickly redistributed inside the octahedron. As a
result a  double-well potential, designed for the vibration of the other oxygen
 ions belonging to the same octahedron, is formed. The longitudinal stretching
phonons with $\omega_{ph}>\omega_0$ are involved in this fast process. After
the relaxation of the deformed surrounding, the pair becomes prepared for the
next hopping, and the described process repeats again and again, providing a
resonant tunneling of the pairs on the large distances along [100] directions.

Note that since the BiO$_6$ and Bi\underline{L}$^2$O$_6$ complexes have
different strengths of the Bi-O bonds, the pair transfer in turn is able to
change both the phase and the wave vectors of the stretching phonons. As the
vibration in the double-well potential is of a breathing-type, the wave vectors
remain unchanged when the octahedra ( which the pair is passing by) vibrates in
the breathing mode.  However a value of $\mathbf q$ can be changed in the case
of the vibrations with the short wave-vector. Due to the dispersion of the
longitudinal stretching mode such a change can cost some energy. Thus to
provide the pair motion in this case, an energy should be transmitted to the
phonons and the dissipation of the pair kinetic energy should occur. Of course,
this leads to the decrease of the BEC critical temperature for the pair. Hence
the more exact estimate for $T_c$ requires the solution of a problem of a
self-consistent preparation of the barrier due to the interaction of a pair
with a phonon subsystem in a process of underbarrier tunneling.  This brings us
all the nice physics of the tunneling with the dissipation
\cite{Caldeira,Zawadovski,Prokofev}. Remind, that in the problem of the
tunneling with the dissipation the shape and the height of a potential barrier
in a two-level system are determined self-consistently taking into account an
interaction of a tunneling particle with a thermal reservoir.  Note that the
 more exact evaluation of $T_c$ will give us an experimentally observed
decrease of the critical temperature as a function of the concentration in the
metallic region $x>0.4$.  This decrease should take place mainly due to the
following facts: (i) a decrease of a bosonic density $n_B$, (ii) an increase of
the width of the barrier $\langle N\rangle$, and maybe the most important,
(iii) the decrease of lattice softening which leads to the increase of the
dissipation of the pair kinetic energy due to the stretching phonons
with $\bbox{\rm q}\ne\bbox{\rm q}_b$. Note also, that at temperatures $T>T_c$ a
bosonic subsystem behaves for the concentrations $x>0,37$ as a normal bosonic
metal with a heavy mass $m_B\sim1/\hat{t}_Ba^2$.

Summarizing the discussion above we point out that the two processes are
important for the superconductivity in Ba$_{1-x}$K$_x$BiO$_3$. The vibrations
of the oxygen ions in the double-well potential provides the mechanism for a
transfer of the local pairs from the one Bose-cluster to the other.  At the
same time the pair motion is strongly affected by the stretching longitudinal
vibrations of the oxygen ions in the octahedra which the pair is passing by. We
suppose that the last process should be taken into account to estimate
correctly the critical temperature.

It is worth to notice, that a similar dispersion of the longitudinal stretching
phonons, which leads to the dominant role of the breathing-type phonons at low
temperatures, has been observed also for the high $T_c$ cuprates
La$_{1.85}$Sr$_{0.15}$CuO$_4$ and YBa$_2$Cu$_3$O$_7$ \cite{Pintschovius}.
Taking into account the recent experimental evidence by M\"uller {\em et al.}
\cite{Mueller} of the coexistence of the small bosonic and fermionic charge
carriers in La$_{2-x}$Sr$_x$CuO$_4$, we suppose also to apply our Fermi-Bose
mixture scenario  to HTSC cuprates. Our paper on this problem is in preparation
\cite{preparation}. Note again that the nature of the pairing itself in the
cuprates is definitely of the electronic origin. However, due to the fact that
underdoped HTSC-materials are close to the phase-separation on AFM and
PM-clusters \cite{Jorgensen,Pomyakushin,Emery}, the lattice here can play an
assistant role again providing a pair tunneling between the superconductive PM
metallic clusters via an insulating AFM-barrier. It can also serve as a
limitation on the estimate of the effective critical temperature for the
Bose-condensation of the pairs in our system.

\section{Conclusion}

In conclusion we briefly summarize the main results.

1. The parent compound BaBiO$_3$ represents a system with the initially
preformed local electron and hole pairs. Every pair is spatially and
energetically localized inside an octahedron volume. The localization energy of
a pair determines the transport activation gap $E_a$. The binding energy of a
pair becomes apparent as the optical gap $E_g=E_b$.

2. The new type of the Fermi-Bose mixture is possibly realized in the
superconductive compositions of Ba$_{1-x}$K$_x$BiO$_3$ for $x>0.37$. The
bosonic bands are responsible for the two-particle normal state conductivity.
The overlap of the empty fermionic band $F$ with an occupied valence band
Bi6$s$O2$p$ provides the insulator-metal phase transition and produces the
Fermi-liquid state. This state shunts to a great extent the normal state
conductivity arising from the two-particle Bose transport.

3. The fermionic band $F'$ connected with the pair destruction does not play
any role in the transport. The excitation energy is high enough to guarantee
against the destruction of bosons (a pair binding energy $E_b\approx0.5$\,eV
for the superconductive compositions).

4. Pair localization energy is absent for $x>0.37$ ($E_a=0$), so the bosonic
and the fermionic subsystems are separated only spatially. The interplay
between them is due to the dynamic exchange
Bi\underline{L}$^2$O$_6\leftrightarrow$BiO$_6$, which causes a free motion of
the local pairs in the real space.

5. The pairing mechanism in the bismuthates is more probably of the electronic
than of the phonon-mediated origin. The existence of the local pairs and their
tunneling between the neighboring octahedra are the reasons for the appearance
of the double-well potential, which describes the vibration of the oxygen ions.
The lattice is involved in the superconductivity more probably by providing the
motion of the local pairs.

Finally, we would like to emphasize that the scenario of the Fermi-Bose mixture
allows us to describe qualitatively an insulator-metal phase transition and a
superconductive state in BKBO in the framework of the one common approach. To
some extent this scenario explains the contradictions observed experimentally
by the UPS and XPS \cite{Namatame2,Qvarford}, the EXAFS and XANES
\cite{Pisma,MenushenP,Salem}, and the Raman \cite{Sugai,Sugai3,Tajima3}
spectroscopies,  as well as by an  inelastic neutron scattering
\cite{Braden2}, the transport and the optical measurements
\cite{Uchida,Hellman1,Blanton}. Nevertheless, the additional experiments are
required to make a definite conclusion about the nature of the
superconductivity in these systems.

First of all we propose two direct experiments to test our model. (i) To
synthesize a new compound Ba$_{1-x}$La$_x$PbO$_3$, which should be
superconductive in accordance with our point of view. (ii) To provide the Raman
scattering experiment of the superconducting Ba$_{0.6}$K$_{0.4}$BiO$_3$
compound using a resonance optical excitation in the range of the optical
pseudogap $E_g\approx0.5$\,eV. In this case the appearance of the additional
Raman modes due to local dynamic distortions should be observed at the
pair destruction in accordance with our model.

Besides, it is important to carry out the more precise measurements of the
specific heat in the bismuthates for $T\sim T_c$. We know that the specific
heat behaves as $C_B\sim(T/T_c)^{3/2}$ for the  temperatures $T<T_c$, and
$C_B={\mathit const}$ for $T\gg T_c$ in a three-dimensional Bose-gas. As a
result, there is a $\lambda$-point behavior of the specific heat for $T\sim
T_c$. However, in the Fermi-Bose mixture there is an additional contribution
from a degenerate Fermi-gas $C_F\sim\gamma T$. This contribution could in
principle destroy a $\lambda$-point behavior of the specific heat in the
Fermi-Bose mixture. Note that the currently available experimental results in
the bismuthates signal a smooth behavior of the specific heat near $T_c$
\cite{Stupp}, because in all the experiments the contributions from the
degenerate Fermi and Bose-gases are masked by a larger lattice contribution.

Another important measurement, which can be proposed to elucidate the nature of
the superconductivity, is a measurement of the thermopower in the normal state
of the bismuthates.  According to the ideas of Larkin {\em et al.}
\cite{Geshkenbein}, the Seebeck coefficient
$S_0=n_B(T)/n_0\ln(n_B(T)/n_0)+\alpha T/\varepsilon_F$ in the Fermi-Bose
mixture is much larger by absolute value than the usual value $S_0=\alpha
T/\varepsilon_F$ in the normal state of an ordinary fermionic metal. Moreover,
the sign of the thermopower can become negative, which is also rather unusual.
Note that up to now there is a lot of controversy in the measurements of
Seebeck coefficient in the bismuthates (see Ref.~\onlinecite{Uher} and
references therein).

\acknowledgements
We acknowledge fruitful discussion with N.\,M.~Plakida, Yu.~Kagan, P.~Fulde,
P.~Woelfe, and A.\,N.~Mitin. This work was supported by Russian Foundation for
Basic Research (Grants 99-02-17343 and 98-02-17077), INTAS grant 97-0963, and
Program ``Superconductivity'' (Grant 99010). M.~Yu.~K is grateful for the
grant of the President of Russia 96-15-96942 and acknowledges a hospitality of
Max-Planck-Institut in Dresden.

%\bibliography{We&K}
%\bibliographystyle{prsty}

\end{document}